# X-RAY POPULATIONS IN GALAXIES

G. Fabbiano

*Harvard-Smithsonian Center for Astrophysics, 60 Garden St., Cambridge MA 02138, USA*

**Abstract:** Today's sensitive, high-resolution *Chandra* X-ray observations allow the study of many populations of X-ray sources. The traditional astronomical tools of photometric diagrams and luminosity functions are now applied to these populations, and provide the means for classifying the X-ray sources and probing their evolution. While overall stellar mass drives the amount of X-ray binaries in old stellar populations, the amount of sources in star forming galaxies is related to the star formation rate. Short-lived, luminous, high mass binaries (HMXBs) dominate these young X-ray populations.

## 1. *Chandra* observations of X-ray binary (XRB) populations

It is well known that the Milky Way hosts both old and young X-ray source populations, reflecting its general stellar make up. In 1978, the *Einstein Observatory*, the first imaging X-ray telescope, opened up the systematic study of the X-ray emission of normal galaxies, and revealed populations of X-ray sources, at least in nearby spiral galaxies (Fabbiano 1989). With *Chandra*'s sub-arcsecond angular resolution, combined with CCD photometric capabilities (Weisskopf *et al.* 2000), the study of normal galaxies in X-rays has taken a revolutionary leap: populations of individual X-ray sources, with luminosities comparable to those of the Galactic X-ray binaries, can be detected at the distance of the Virgo Cluster and beyond.

We can now study these X-ray populations in galaxies of all morphological types, down to typical limiting luminosities in the $10^{37}$ ergs s$^{-1}$ range. At these luminosities, the old population X-ray sources are accreting neutron star or black-hole binaries with a low-mass stellar companion, the LMXBs (life-times $\sim 10^{8-9}$ yrs). The young population X-ray sources, in the same luminosity range, are dominated by neutron star or black hole binaries with a massive stellar companion, the HMXBs (life-times $\sim 10^{6-7}$ yrs; see Verbunt & van den Heuvel 1995 for a review on the formation and evolution of X-ray binaries), although a few young supernova remnants (SNRs) may also be expected. At lower luminosities, reachable with *Chandra* in Local Group galaxies, Galactic sources include accreting white dwarfs and more evolved SNRs. Fig. 1 shows two typical observations of galaxies with *Chandra*: the spiral M83 (Soria & Wu 2003) and the elliptical NGC4697 (Sarazin, Irwin & Bregman 2000), both observed with the ACIS CCD detector. The images are color coded to indicate the energy of the detected photons (red 0.3-1 keV, green 1-2 keV and blue 2-8 keV). Populations of point-like sources are easily detected above a generally cooler diffuse emission from the hot interstellar medium. Note that luminous X-ray sources are relatively sparse by comparison with the underlying stellar population, and can be detected individually with the *Chandra* sub-arcsecond resolution, with the exception of those in crowded circum-nuclear regions.

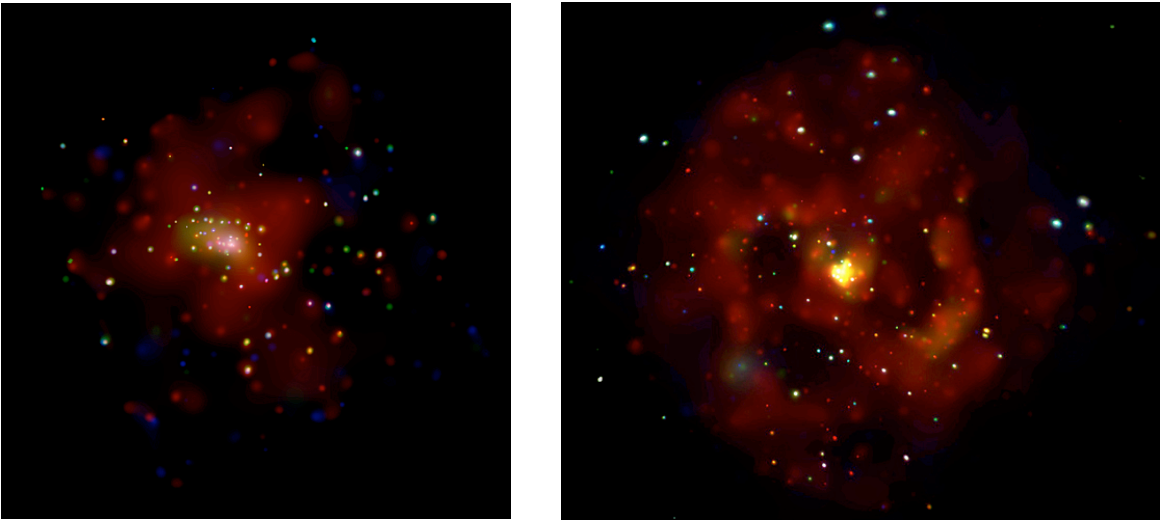

Fig.1 – Chandra ACIS images of NGC4697 (left, box is 8.64 × 8.88 arcmin) and M83 (right, box is 8.57 × 8.86 arcmin). See text for details. Both images are from the *Chandra* web page http://chandra.harvard.edu/photo/; credit NASA/CXC).

To analyze this wealth of data two principal approaches have been taken: (1) a photometric approach, consisting of X-ray color-color diagrams and color-luminosity diagrams, and (2) X-ray luminosity functions (XLFs). Whenever the data allow it, time and spectral variability studies have been pursued. Optical and radio identifications of X-ray sources and association of their position with different galaxian components are also being increasingly undertaken.

**2. X-ray colors**

The use of X-ray colors to classify X-ray sources is not new. For example, White & Marshall (1984) used this approach to classify Galactic XRBs, and Kim, Fabbiano & Trinchieri (1992) used *Einstein* X-ray colors to study the integrated X-ray emission of galaxies. Unfortunately, given the lack of standard X-ray photometry to date, different definitions of X-ray colors have been used in different works; in the absence of instrument corrections, these colors can only be used for comparing data obtained with the same observational set up. Colors, however, have the advantage of providing a spectral classification tool when a limited number of photons are detected from a given source, which is certainly the case for most X-ray population studies in galaxies. Also, compared with the traditional derivation of spectral parameter via model fitting, color-color diagrams provide a relatively assumption-free comparison tool. *Chandra*-based examples of this approach can be found in Zezas *et al.* 2002a, b and Prestwich *et al.* 2003, among others. The X-ray color-color diagram of Prestwich *et al.* 2003 (fig. 2) illustrates how colors offer a way to discriminate among different types of possible X-ray sources.

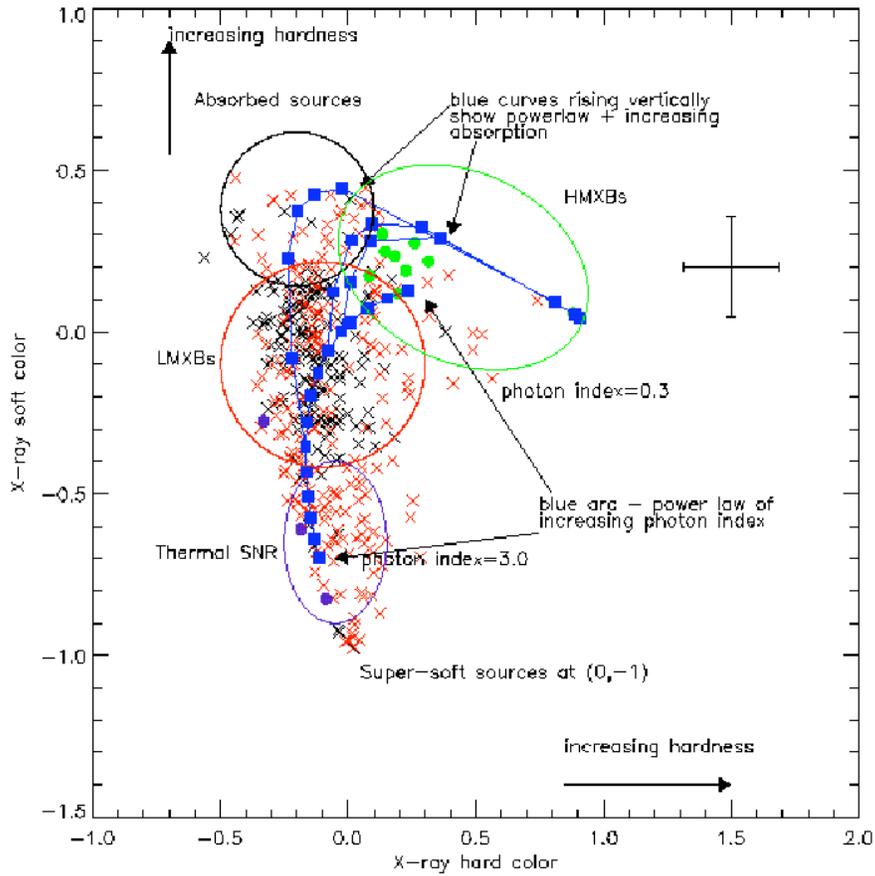

Fig. 2 – *Chandra* color-color diagram from Prestwich *et al.* 2003

## 3. XLFs and parent stellar populations

Luminosity functions are well known tools in observational astrophysics. XLFs have been used to characterize different X-ray binary populations in the Milky Way (e.g., Grimm, Gilfanov & Sunyaev 2002), but these studies have always required a model of the spatial distribution of the sources, so to estimate their luminosities, which is inherently a source of uncertainty. External galaxies, instead, provide clean source samples, all at the same distance. Moreover, the detection of X-ray source populations in a wide range of different galaxies allows us to explore global population differences that may be connected with the age and or metallicity of the parent stellar populations. XLFs establish the observational basis of X-ray population synthesis (Belczinsky *et al.* 2004).

The XLFs have been fitted with power laws or broken power laws. The main parameters are: power-law slope (giving the relative luminosity distribution of X-ray sources), normalization (the total number of sources) and eventual breaks, pointing to changes in the X-ray source population (for example, different breaks are seen in the XLFs of the inner and outer bulge of M31, Kong *et al.* 2002). *Chandra* and XMM-Newton studies of M31 have revealed a variety of XLFs, connected with the different stellar populations of the field in question (see review of Fabbiano & White 2005 and references therein; Kong *et al.* 2003). In M81, the XLF of the spiral arm stellar population is flatter than that of the inter-arm and bulge regions, consistent with the prevalence of short-lived luminous HMXBs in younger stellar populations (Tennant *et al.* 2001, fig. 3; Swartz *et al.* 2002).

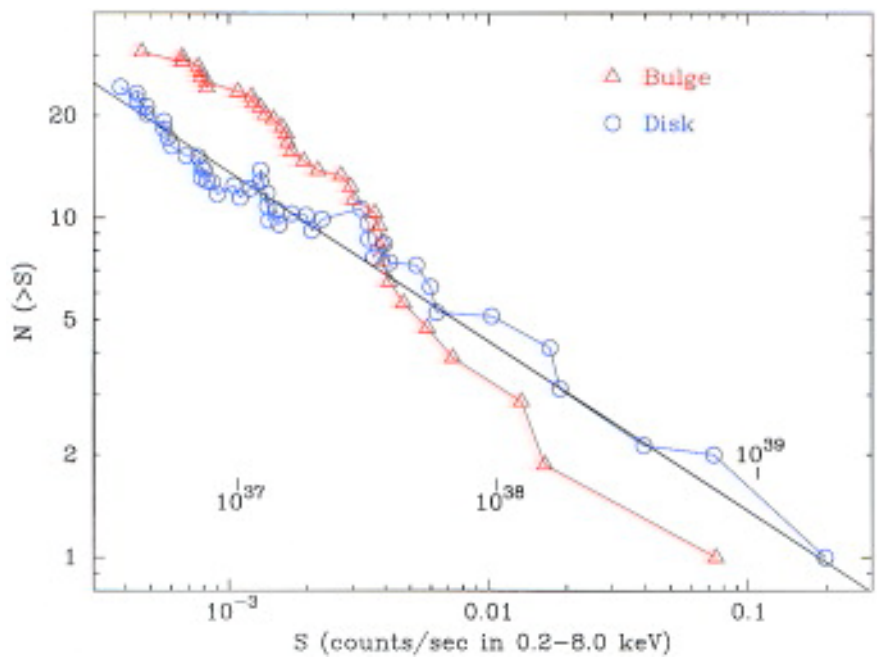

Fig. 3 – Bulge and disk XLFs of M81 (Tennant *et al.* 2001).

In general, flatter XLFs are found in more actively star-forming galaxies (e.g., a cumulative slope of ~ -0.5 is found in the actively star-forming merger system, the Antennae; Zezas & Fabbiano 2002). The early comparisons of XLFs of different types of galaxies (Zezas & Fabbiano 2002; Kilgard *et al.* 2002) also suggested that the normalization are related to either the star formation rate (SFR) or the mass of the parent galaxy. Grimm, Gilfanov & Sunyaev (2003) took these ideas a step further, suggesting that HMXB XLFs follow a universal -0.6 cumulative power-law, with normalization proportional to the SRF. Gilfanov (2004) suggests that the normalization of LMXB XLFs is driven by the stellar mass of the galaxy (see also Kim & Fabbiano 2004).

In E and S0 galaxies the shape of the XLF has also been parameterized with models consisting of power-laws or broken power-laws. The overall shape (in a single power-law approximation in the observed range of ~ 7 x $10^{37}$ to a few $10^{39}$ ergs s$^{-1}$) is fairly steep (cumulative slopes -1 or steeper), i.e. with a relative dearth of high luminosity sources, when compared with the XLFs of star-forming galaxies. A lot of discussion has focused on a reported break at ~2-5 x $10^{38}$ ergs s$^{-1}$, near the Eddington limit of an accreting neutron star (Sarazin, Irwin & Bregman 2000 in NGC4697), which may be related to the transition between neutron star and black hole binaries in the population. Although some reported breaks are the result of incompleteness at the low luminosities (Kim & Fabbiano 2003), and typically breaks are not found in the completeness corrected XLFs of individual galaxies, Kim & Fabbiano (2004, see also Gilfanov 2004) report a break at $(5 \pm 1.6) \times 10^{38}$ ergs s$^{-1}$ in the co-added corrected luminosity function for a sample of 14 E and S0 galaxies (fig. 4).

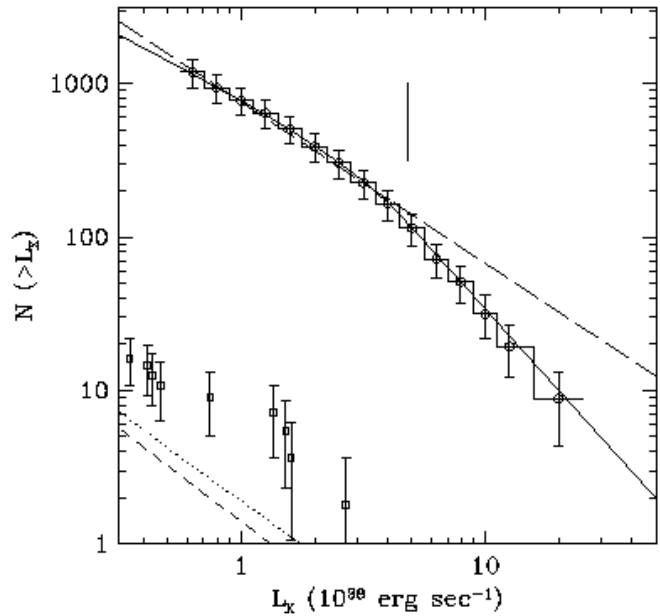

Fig. 4 - Co-added, completeness corrected XLF of 14 E and S0 galaxies, compared with the LMXB XLF of the Galaxy and M31 (Kim & Fabbiano 2004

The paucity of very luminous X-ray sources in galaxies makes uncertain the definition of the high luminosity XLF, which may be better approached by co-adding `consistent' samples of X-ray sources (e.g., Kim & Fabbiano 2004), but still uncertainties persist. Interestingly, the evaluation of the total X-ray luminosity of a galaxy may be significantly affected by statistics when a relatively small number of X-ray sources are detected (Gilfanov, Grimm & Sunyaev 2004).

Compact X-ray sources are notorious for their variability and this variability could in principle also affect the XLF, which is typically derived from a snapshot of a given galaxy. However, repeated *Chandra* observations in the case of both M33 (Grimm *et al.* 2005) and the Antennae galaxies (Zezas *et al.* 2005, in preparation) demonstrate that the XLF is remarkably steady, even when high luminosity sources clearly vary.

**4. Conclusions**

The results discussed in this talk represent only the beginning of what I hope will be a very fruitful field of investigation for many years to come. The tools that are being developed for characterizing and understanding the X-ray source populations of nearby galaxies lay the foundation of future work in X-ray population synthesis (see Belczynski et al 2004). This approach, and future more sensitive high resolution X-ray observations, such as those anticipated from a possible *Gen-X* mission now in preliminary study by NASA, will allow the study of the X-ray evolution of galaxies. Given the strong link between young X-ray sources and star formation, these studies will also provide a direct probe of the process of galaxy formation and evolution.

This work was partially supported under NASA contract NAS8-39073 (CXC).
This material was also covered in a review talk delivered at the COSPAR Colloquium "Spectra and Timing of Compact X-ray Binaries", held in Mumbai (India), January 17-20, 2005.